\documentclass[conference]{IEEEtran}
\pdfoutput=1
\IEEEoverridecommandlockouts
\usepackage{cite}
\usepackage{amsmath,amssymb,amsfonts}
\usepackage{algorithmic}
\usepackage{graphicx}
\usepackage{textcomp}
\usepackage{xcolor}
\def\BibTeX{{\rm B\kern-.05em{\sc i\kern-.025em b}\kern-.08em
    T\kern-.1667em\lower.7ex\hbox{E}\kern-.125emX}}
    
\usepackage{booktabs} 
\usepackage{color,soul}
\usepackage{booktabs,array,booktabs,pgfplots,
			tikz}
			
\ifCLASSOPTIONcompsoc
\usepackage[caption=false,font=normalsize,labelfont=sf,textfont=sf]{subfig}
\else
\usepackage[caption=false,font=footnotesize]{subfig}
\fi

\usepackage{listings}
\usepackage{syntax} 
\usepackage[inline]{enumitem}
\usepackage{amsthm}
\usepackage{pgfplotstable}
\usepackage{bm}
\usepackage{tikz-uml}
\usepackage[english]{babel}

\usetikzlibrary{petri}
\usetikzlibrary{positioning,automata}
\usetikzlibrary{arrows.meta}

\pgfplotsset{compat=1.3}

\newtheorem{definition}{Definition}

\begin{document}

\title{TiLA: Twin-in-the-Loop Architecture for Cyber-Physical Production Systems
\thanks{This work was supported by Delta-NTU Corporate Lab for Cyber-Physical Systems with funding support from Delta Electronics Inc. and the National Research Foundation (NRF) Singapore under the Corp Lab@University Scheme.}
}

\author{\IEEEauthorblockN{Heejong Park, Arvind Easwaran}
\IEEEauthorblockA{Nanyang Technological University, Singapore \\
\{hj.park, arvinde\}@ntu.edu.sg}
\and
\IEEEauthorblockN{Sidharta Andalam}
\IEEEauthorblockA{Delta Electronics Inc., Singapore \\
sidns5@gmail.com}
}

\maketitle

\begin{abstract}
Digital twin is a virtual replica of a real-world object that lives simultaneously with its physical counterpart. Since its first introduction in 2003 by Grieves, digital twin has gained momentum in a wide range of applications such as industrial manufacturing, automotive and artificial intelligence. However, many digital-twin-related approaches, found in industries as well as literature, mainly focus on modelling individual physical things with high-fidelity methods with limited scalability. In this paper, we introduce a digital-twin architecture called TiLA (\textbf{T}win-\textbf{i}n-the-\textbf{L}oop \textbf{A}rchitecture). TiLA employs heterogeneous models and online data to create a digital twin, which follows a Globally Asynchronous Locally Synchronous (GALS) model of computation. It facilitates the creation of a scalable digital twin with different levels of modelling abstraction as well as giving GALS formalism for execution strategy. Furthermore, TiLA provides facilities to develop applications around the twin as well as an interface to synchronise the twin with the physical system through an industrial communication protocol. A digital twin for a manufacturing line has been developed as a case study using TiLA. It demonstrates the use of digital twin models together with online data for monitoring and analysing failures in the physical system.
\end{abstract}

\section{Introduction}

The era of Industry 4.0 brings increased connectivity among devices, predictability through a large volume of data, and an ability to accurately capture states of the manufacturing shop-floor. Since the introduction of a digital twin concept by Grieves~\cite{grieves2017digital}, it has been considered as one of the key enabling technologies that pushes boundaries of factory digitalisation. Digital twin is a realistic virtual representation of machines or any form of living or non-living physical things that can accurately monitor, predict and optimise their operations. An advancement of today's information and communication technologies (ICT), low-powered sensor and actuator devices together with intelligent software and hardware platforms, make it possible to create a digital twin that tightly interconnects cyber and physical spaces. 

Nevertheless, many works found in today's literature on digital twin lack support for integration of heterogeneous models with online data for creating a scalable digital twin. In our view, a key to success in implementing a digital twin is to identify the requirements of the target application and provide appropriate models that can accurately capture the characteristics of the physical system. To achieve this, a digital twin architecture should support the utilisation of different modelling strategies for creating the twin, where each model specialises in certain application domains. Furthermore, it is essential to support a method to combine such heterogeneous models in a consistent manner and utilise gathered online data from a physical system to provide a holistic understanding of the physical system’s state.

One of the common mistakes in the digital twin implementation is the attempts to provide every detail about its physical twin through a model, while not every information is useful nor utilised in the application. In addition, it is not always feasible to model every part of the physical phenomena using detailed modelling mechanisms, especially for large scale systems. Building and deploying a digital twin in the real world scenario requires not only its ability to faithfully capture dynamics of physical systems using high-fidelity techniques, but also to combine different levels of model abstractions in a deterministic way for efficiency and scalability.

Recently digital twin has received considerable attention in the fields of manufacturing, process plants, cloud computing, etc~\cite{negri2017review,kritzinger2018digital}. Authors in \cite{tao2017digital} presented a digital twin architecture that consists of a service system, containing data and algorithms, and a virtual layer that provides high fidelity models of the physical system. Nevertheless, methods for interconnecting different model abstractions and modelling formalisms are missing in their work. A digital twin is used in a software-defined control (SDC) framework in \cite{lopez2018software}. Their architecture, however, assumes the existence of an MES where the twin does not directly interact with the physical system and an online use-case of the twin is missing. A cloud-based digital twin architecture reference model is introduced in \cite{alam2017c2ps}. On the high-level, the model of cyber things are composed as a set of finite state machines. However, unlike our twin model, their model composition strategy does not support asynchrony which is desirable for scalability. Tools such as Ptolemy~II~\cite{ptolemaeus2014system}, and INTO-CPS~\cite{larsen2016integrated} are more focused on the simulation of cyber-physical systems (CPS) using different levels of formal models of computation (MoC). Yet, their focus have been on the offline simulation of cyber-physical systems, whereas the digital twin evolves with the system in real-time.

To address the aforementioned issues and challenges, this paper introduces a digital twin architecture TiLA (\textbf{T}win-\textbf{i}n-the-\textbf{L}oop \textbf{A}rchitecture) that merges benefits of the traditional model-based design and data-based modelling approaches for creating and deploying digital twins for Cyber-Physical Production Systems (CPPS). Here, all physical operations of the production system are captured and accessed via digital twin in the cyber space. In particular, our architecture enables building of a digital twin through the composition of different levels of modelling abstractions by using the formal Globally Asynchronous Locally Synchronous (GALS) model of computation (MoC)~\cite{malik2010systemj,jebali2014grl,berry2001multiclock}. The contributions of this paper are:
\begin{enumerate}
	\item A digital twin architecture that employs heterogeneous modelling techniques to capture the digital twin's behaviour. Models with different abstractions execute and communicate with each other based on GALS MoC.
	\item Support for a cyber-physical infrastructure that facilitates a single source of truth for the applications that depend on the online status of the physical system consisting of various assets and provides an accurate view of the physical system through the digital twin. 
	\item A digital twin case-study which demonstrates the use of heterogeneous models and online data for finding the root-causes of failure in a manufacturing line. 
	The result also showed that the use of heterogeneous models is favourable for maintaining a performance of a digital twin that has to match with a speed of the physical system.
\end{enumerate}

The rest of this paper is organised as follows: 
Section~\ref{sec:arch} introduces the three-layered software architecture of the TiLA. The semantics of GALS MoC for the digital twin used in the architecture is given in Section~\ref{sec:gals}. 
Section~\ref{sec:case} presents a digital twin case study called fault monitoring and analysis for a factory assembly consisting of a two-link planar robot, conveyor belts, and inspection station for analysis. Finally, conclusions and potential future works are presented in Section~\ref{sec:conclusion}.

\section{Architecture}\label{sec:arch}

\begin{figure}[t!]
	\centering
	\includegraphics[page=1,width=\linewidth]{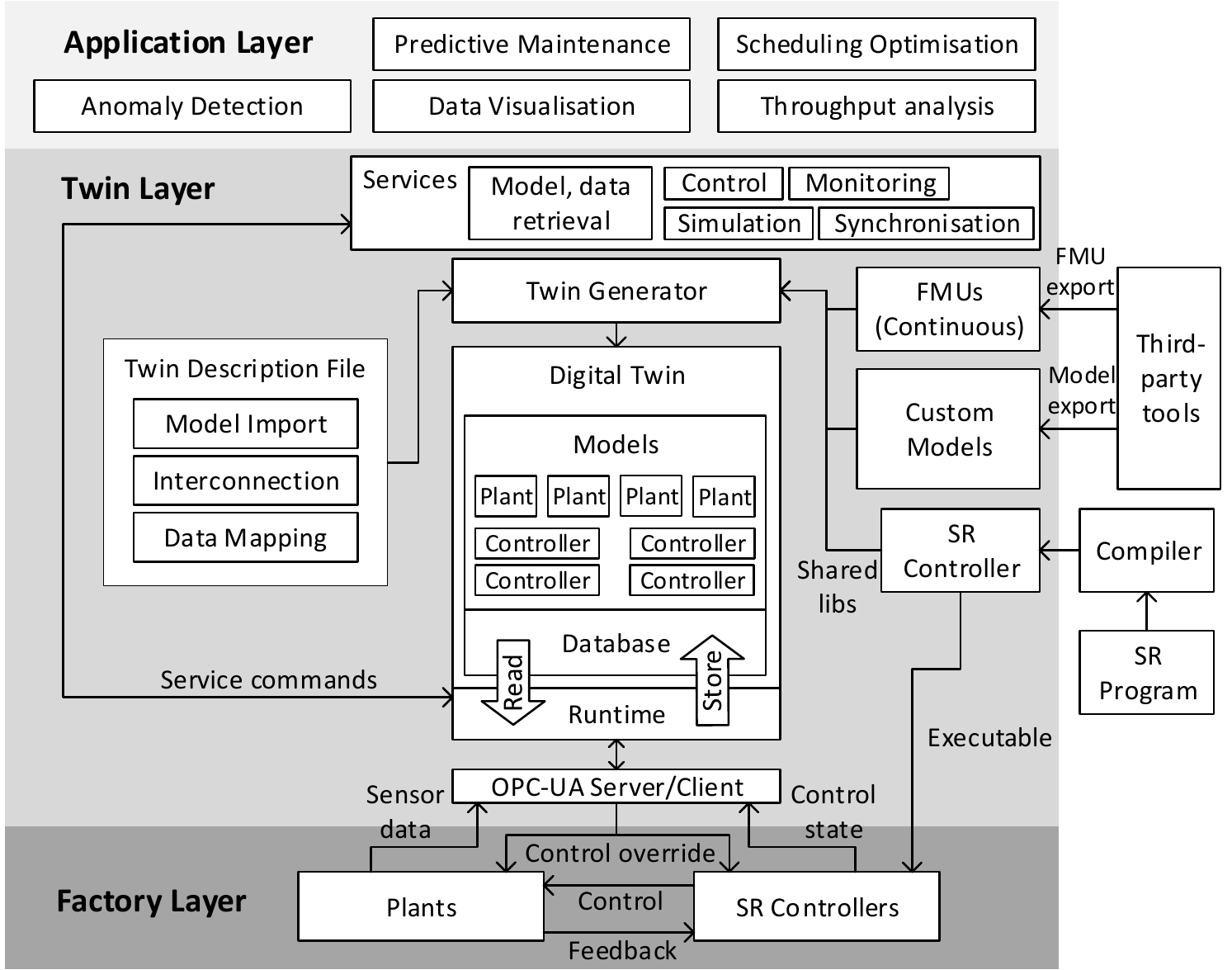}
	\caption{An architecture of TiLA: Twin-in-the-Loop Architecture}
	\label{fig:title}
\end{figure}

This section presents the three-layer architecture of TiLA. An overview of the architecture is depicted in Fig.~\ref{fig:title}. Each layer has a number of facilities that help the higher-level layers to perform more concrete tasks.

\subsection{Factory Layer}

The Factory Layer controls the manufacturing process and forwards data to one level above called the Twin Layer. The data are either sensor values collected from the physical plant, such as temperature, battery levels, object positions, etc., or control related data such as outputs of the controllers and their internal state information. The Factory Layer can also receive control commands from the Twin Layer that override the local control actions or tune the parameters of the control logic. Controllers need not know how digital twins are maintained nor how gathered data are utilised. 

As shown in Fig.~\ref{fig:title}, designers model control-logic using synchronous reactive (SR) languages such as Esterel~\cite{berry1992esterel} or their variants, which are well suited for capturing the concurrent and preemptive nature of control-dominated behaviours in a deterministic way. The SR model is based on a formal mathematical foundation which makes the program amenable to efficient execution and verification.

\subsection{Twin Layer}\label{sec:twin-layer}

\subsubsection{Digital Twin}\label{sec:dt}

A digital twin in TiLA is composed of plant and controller models, which are interconnected through uni-directional signals, and a collection of data models for organising sensor as well as control data gathered from the physical assets.

\begin{definition}[Digital twin]\label{def:dt}
	A digital twin $DT$ in TiLA is a tuple $\langle \mathcal{C},\mathcal{T},\mathcal{S},\mathcal{E}_q,\mathcal{E}_s,\mathcal{D} \rangle$, where
	\begin{itemize}
		\item A finite set of clock-domains $\mathcal{C}$ such that $\ \forall \mathcal{C}_i \in \mathcal{C},\ \mathcal{C}_i$ is a set of models $\{m_1,\dots,m_n\} $ and $\ \forall \mathcal{C}_i,\mathcal{C}_j \in \mathcal{C},\ i\neq j,\ \mathcal{C}_i\cap \mathcal{C}_j = \emptyset $.
		\item A set of sequences of ticks $ \mathcal{T} $ such that $\forall \mathcal{T}_i \in \mathcal{T},\ \mathcal{T}_i=\{(n,r) \mid n\in\mathbb{N}, r\in\mathbb{R},n\geq 0 \wedge r\geq 0\}$ for a clock-domain $ \mathcal{C}_i $, and $ n $ and $ r $ indicate a logical time instant and its corresponding value in the continuous time domain, respectively.
		\item A finite set of set of signals $\mathcal{S}$ such that $ \mathcal{S}_i \in \mathcal{S} $ is a set of signals for a clock-domain $ \mathcal{C}_i $. Each signal $ s\in \mathcal{S}_i $ consists of a status and a value pair $(s_{st},s_{v})$ where $s_{st}\in\{0,1,\bot\},\ s_v\in \mathbb{R}$.
		\item A finite set of connections between two models in different clock-domains $\mathcal{E}_q \subseteq \mathcal{C}_i\times \mathcal{S}_i\times \mathcal{S}_j\times \mathcal{C}_j$.
		\item A finite set of connections $\mathcal{E}_s \subseteq \mathcal{C}_i\times \mathcal{S}_i\times \mathcal{C}_i $ between models within the same clock-domain.
		\item A collection of data models $\mathcal{D}\subseteq \sum^\ast\times\mathbb{R}^k$ where $ \sum^\ast $ is a set of finite-length words over alphabets.
	\end{itemize}
\end{definition}

\begin{figure}[t!]
	\centering
	\subfloat[]{
		\includegraphics[page=1,scale=0.42]{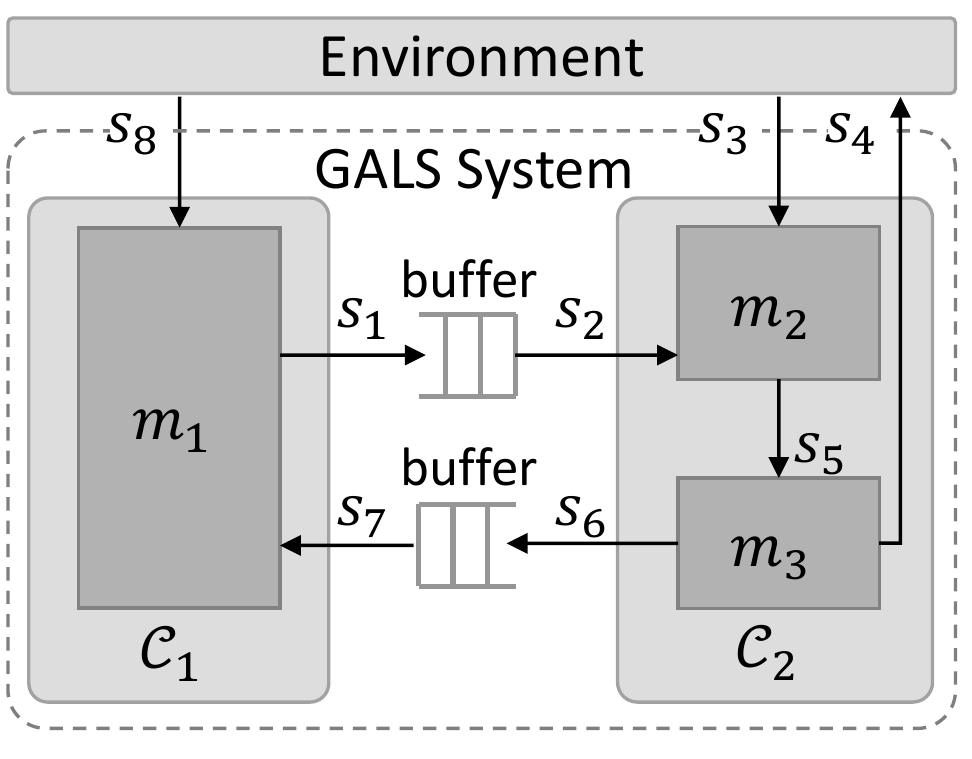}
		\label{fig:gals}}%
	\hspace*{\fill}
	\subfloat[]{
		\includegraphics[page=2,scale=0.44]{figures/slides.pdf}
		\label{fig:ticks}}
	\caption{A graphical illustration of \protect\subref{fig:gals} a GALS system for a digital twin in TiLA and \protect\subref{fig:ticks} corresponding tick execution trace for $ \mathcal{C}_1 $ and $ \mathcal{C}_2 $}
	\label{fig:gals-illustration}
\end{figure}

Intuitively, plants and controllers in our digital twin are heterogeneous models following GALS MoC~\cite{malik2010systemj,jebali2014grl,berry2001multiclock} on a top level. For example, Fig.~\ref{fig:gals} shows a graphical overview of a GALS system consisting of two clock-domains $ \mathcal{C}_1 $ and $ \mathcal{C}_2 $. $ \mathcal{C}_1 $ has a single model $ m_1 $, whereas $ \mathcal{C}_2 $ has two models $ m_2 $ and $ m_3 $. All models in the same clock-domain run concurrently in lock-step fashion and synchronise with each other at a sequence of logical instants called \textit{ticks}. As shown in Fig.~\ref{fig:ticks}, each clock-domain has its own sequence of ticks, for example $ \mathcal{T}_1 =\{t_1,t_2,t_3,\ldots\}$ and $ \mathcal{T}_2 =\{t_1,t_2,t_3,\ldots\}$ fed to $ \mathcal{C}_1 $ and $ \mathcal{C}_2 $, respectively. The arrows with the vertical bar indicate the tick instants.
Execution of a tick consists of three main steps: (1) at the beginning of a tick (BOT) $ t \in \mathcal{T}_i $, inputs to a clock-domain are captured from the external environment and transformed into a logical entity called \textit{signals}. Each signal consists of a ternary status 1, 0, or $ \bot $ (unknown) and possibly a value. For each clock-domain tick, all signal statuses are resolved from unknowns to 0 or 1. In Fig.~\ref{fig:ticks}, there is one instant of tick for $ \mathcal{C}_2 $, depicted by a dotted blue box, where the clock-domain captures input signal $ s_3=a $ at BOT. (2) $ \mathcal{C}_2 $ then makes internal computations and emits $ s_4=b $. (3) As a result, $ b $ is transmitted to the external environment at the end of a tick (EOT). These steps take place at the same tick instant and are assumed to be \textit{instantaneous} with respect to the speed of the environment~\cite{berry1992esterel}.

Signal is associated with a particular clock-domain's tick. Since ticks for different clock-domains are unrelated (not synchronised), a signal event generated by one clock-domain cannot be captured by another clock-domain in a reliable way. Therefore in TiLA, a point-to-point First-in-First-Out (FIFO) buffer is used for exchanging data between models in different clock-domains; see connections between models via buffers in Fig.~\ref{fig:gals}. 
Finally, TiLA contains a set of data models for storing and analysing online data gathered from the physical twin. Each digital twin model $ m \in \mathcal{C}_i $ has an associated data model $ \mathcal{D}_{m} \subseteq \mathcal{D} $ that can be utilised by the applications using the digital twin for monitoring and analysing the status of the physical system. It is structured in a similar fashion as the relational database model where the signal values are stored as elements in the rows of a table with additional information attached as string literals. For example, when the twin model needs to be synchronised with the online data with a specific timeline, the data model can provide such information.

\subsubsection{Twin Description File (TDF) and Models}

This configuration file for a digital twin is one of the inputs to our architecture which contains the following components:
\begin{enumerate*}
	\item \textit{Model import} describes a list of models used to create a digital twin.
	\item \textit{Interconnection} describes input and output interconnections between the models.
	\item \textit{Data mapping} describes relations between the data model and the digital twin.
\end{enumerate*}
A TDF together with models exported from third-party tools are used to build a digital twin by the Twin Generator. White-box models such as SR programs, Petri-Nets \cite{peterson1981petri} and finite state machines can be imported together with black-box models encapsulated in Functional Mock-up Unit (FMU). FMU is one of the Functional Mock-up Interface (FMI) standards~\cite{blochwitz2011functional} that provides a set of APIs for external tools to interact with the FMUs. Once models are imported for creating a twin, TiLA's runtime environment implements the master algorithm based on the GALS MoC, which is not part of the FMI standard and needs to be implemented by a tool that uses the FMUs. To include non-FMU based models exported by third-party tools, the architecture provides a minimal set of wrappers to access these models. As a result, TiLA integrates both FMU and non-FMU based models to create a digital twin based on the GALS MoC.

\subsection{OPC-UA Interface}

\begin{figure}[t!]
\centering
\includegraphics[page=6,width=0.85\linewidth]{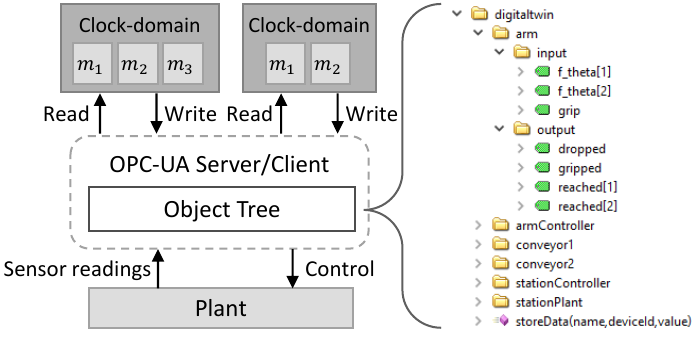}
\caption{An interface between a digital twin and a physical plant via the OPC-UA object model.}
\label{fig:opcua-object}
\end{figure}

The Twin Layer collects measurable sensor and control data from the Factory Layer through a cyber-physical interface. Currently, this interface is implemented using the OPC-UA architecture~\cite{web:opc-ua}, which includes a machine-to-machine communication protocol for industrial automation. In the Twin Layer, the OPC-UA server provides a list of objects that can be accessed by the clients (physical plant and controllers). A notion of object is similar to objects in object-oriented programming. OPC-UA clients operate on a set of variables and methods created by the server which are based on the information provided in the TDF by a designer. A typical OPC-UA object tree created from the TDF is shown in Fig.~\ref{fig:opcua-object}, which bridges between digital twin models and their physical counterpart. Physical plants and SR controllers can write or read I/O signals of a model via the OPC-UA objects under the \texttt{input} and the \texttt{output} folders. These objects are kept updated with the latest signal values of the models at the end of respective clock-domain ticks. When applications request an override operation and writes a value to input signal of a model, such value is written to the associated variable in the OPC-UA tree and also forwarded to the physical plant as a control signal via an OPC-UA client. Sensor data that is not directly associated to any model inputs can be stored into database via an OPC-UA method \texttt{storeData()}.

\subsubsection{Services}

Services in the Twin Layer are used by applications for retrieving and delivering information from/to the digital twin as well as for enhancing the twin's functionality with additional features. They allow applications to utilise different modes of operations of the digital twin as mentioned in Section~\ref{sec:twin-layer}. For example, an application can request a TDF for visualising the current system layout via the \textit{model and data retrieval} service, which also provides a way to access the digital twin database in a secure and reliable way. The \textit{simulation service} is used for requesting a simulation with a specific twin configuration. The \textit{control service} is used by authenticated applications to override control of physical plants for remote maintenance and reconfiguration. The \textit{monitoring} service can be configured to capture various events, such as sensor data exceeding a threshold value and discovering certain patterns from machine operations, based on data gathered both from the twin as well as the physical plant. These events are transformed into domain-specific knowledge in the Application Layer. Lastly, the \textit{synchronisation} service allows the digital twin to synchronise with the physical plant by incorporating online data during the real-time execution of the twin models.

\subsection{Application Layer}

\begin{table}[t!]
	\centering
	\caption{APIs for applications to interact with the digital twin}
	\label{tab:restful-api}
	\setlength{\tabcolsep}{2.5pt}
	\scriptsize
	\begin{tabular}{llllm{3.0cm}}
		\toprule
		\textbf{Method} & \textbf{Resource} & \textbf{Data}& \textbf{Returns}  &\textbf{Description}\\
		\midrule
		GET & \texttt{/twins} &-& TDF & Returns description of the digital twin in TDF\\
		POST & \texttt{/simulate}& TDF &JSON& Requests digital twin simulation\\
		GET & \texttt{/map} &-&AML& Returns engineering data of a digital twin in AutomationML\\ 
		GET & \texttt{/map/collada/fn}&-&.dae& Returns geometric data in COLLADA format\\
		POST & \texttt{/create/observer} &LTL&JSON& Creates an observer for monitoring events\\
		GET & \texttt{/data/query} &SQL&JSON& Retrieves data from database\\ 
		POST/GET & \texttt{/model/name/port} &JSON&JSON& Writes or reads signals of the digital twin model\\
		\bottomrule
	\end{tabular}
\end{table}

\begin{figure}
\centering
\resizebox{0.75\linewidth}{!}{
\begin{tikzpicture}
\begin{umlseqdiag}
\umlobject[no ddots] {app}
\umlobject[x=5.6,no ddots] {service}
\begin{umlcall}[op={\texttt{GET:/twins}},return=file:TDF]{app}{service}
\end{umlcall}
\begin{umlcall}[padding=-1,dt=5,op=\texttt{\shortstack{POST:/create/observer\\Content: spec}},return=\parbox{3.8cm}{\texttt{\{"topic": "event-1", "uri": "..", ...\}}}]{app}{service}
	\begin{umlcallself}[op=\texttt{createTopic()}]{service}
	\end{umlcallself}
\end{umlcall}
\begin{umlcallself}[op=\texttt{subscribe("event-1")}]{app}
\end{umlcallself}
\begin{umlcallself}[op=\texttt{execModels()}]{service}
\end{umlcallself}
\begin{umlcall}[type=asynchron, op=\texttt{publishEvent()},padding=-3]{service}{app}
	\begin{umlcallself}[op=\texttt{callback()}]{app}
	\end{umlcallself}
\end{umlcall}
\end{umlseqdiag}
\end{tikzpicture}
}
\caption{A use case of the monitoring service for digital twin applications}
\label{fig:case-seq}
\end{figure}
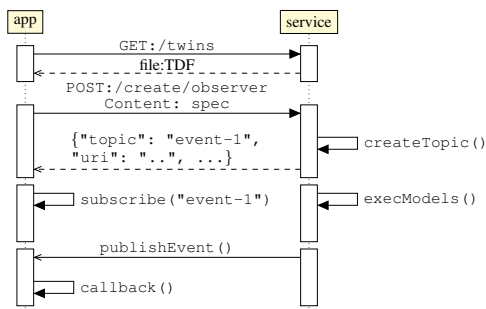

Applications in this layer interact with the digital twin for different use-case scenarios such as throughput monitoring, predictive maintenance, scheduling optimisation, etc. It is assumed that every application-specific logic is implemented in this layer and the digital twin only provides ``facts'' about its physical counterpart rather than trying to interpret what they mean to different use-cases. For instance, the Twin Layer does not need to know if a particular twin model is describing mechanical or electrical components nor if certain events detected from a physical plant are anomalous or expected, or completion of a manufacturing process. Interpretation of such information is done in the Application Layer that makes the digital twin generic and usable in various application domains rather than tailoring each digital twin for specific use-cases.

A use-case of the digital twin for a monitoring application is illustrated in a sequential diagram as shown in Fig.~\ref{fig:case-seq}. In this scenario, the application first requests an information on the currently available digital twin models via an HTTP GET request \texttt{/twin} to the monitoring service, which returns a TDF as a response. After examining the returned file, an HTTP POST \texttt{/create/observer} is sent with a monitoring specification encoded in Linear Temporal Logic (LTL)~\cite{pnueli1981temporal}. The monitoring service then generates an observer according to the received LTL formula and creates a topic that can be subscribed by the application. When the observer detects the monitoring event, the service publish a message to the topic that can be used by the application in a registered callback function.

A list of currently available APIs to access digital twin is shown in Table~\ref{tab:restful-api}. In addition to the APIs used in the previous example, \texttt{/map} returns an engineering data encoded in AutomationML~\cite{drath2008automationml}, such as geometry (e.g. 3D information) and topology of the physical system. Using this data, the application can retrieve more detailed information of the physical system that might not be found in the TDF file. In addition, applications can request offline simulation of the digital twin model via \texttt{/simulate} where different model configurations are passed as a TDF file. The result of simulation is populated in a JSON format and returned to the application upon the completion. Lastly, latest values of input and output signals of individual model can be accessed via POST/GET calls to \texttt{/model/name/port}.

\section{GALS Semantics for the Digital Twin}\label{sec:gals}

GALS is a superset of the SR model where multiple synchronous islands are grouped into clock-domains, and each clock-domain is triggered with its own sequence of ticks. In the following, a notation $ \langle x,\mathcal{S}_i \rangle$ is used to describe the states of $ x $ and corresponding clock-domain signals $ \mathcal{S}_i $, respectively, where $ x $ is $ \mathcal{C}_i $ for a clock-domain and $ m $ for a single model. A clock-domain is a single model or collection of models composed with the synchronous parallel operator $ || $. We adopt a common notation used in~\cite{plotkin2004structural} to describe a state transition of $ x $ where $ \langle x,\mathcal{S}_i \rangle \to \langle x',\mathcal{S}_i' \rangle $ and $ \langle x,\mathcal{S}_i \rangle \to  x'  $ denote a \textit{micro-step} and a \textit{macro-step} state transition of $ x $, respectively. A clock-domain makes a micro-step state transition without progressing its time, therefore this is considered as an intermediate state transition. A macro-step transition occurs when the clock-domain progresses time, i.e. $ t_{i+1} > t_i \iff t^n_{i+1} > t^n_i\wedge t^r_{i+1} > t^r_i$ where $ t^n $ and $ t^r $ are the logical and the physical time components of $ t $. Synchronous composition and execution of models within the same clock-domain $ \mathcal{C}_i $ are then defined as:
\begin{equation}\label{eq:sync}
\frac{\langle m_1,\mathcal{S}_i \rangle \to m_1' ,\  \langle m_2,\mathcal{S}_i \rangle \to   m'_2 }{\langle m_1 || m_2, \mathcal{S}_i \rangle \to m_1'|| m_2',\ t_{m_1'}=t_{m_2'}}
\end{equation}
where $||$ indicates the synchronous parallel operator that ensures model synchronisation at every discrete instant of $ t \in \mathcal{T}_i $ within the same clock-domain. 

As a superset of the SR MoC, local synchronous threads in GALS react to external events in \textit{zero} time with respect to their clock-domain's tick. This implies input and output relationships between models can create cyclic dependencies that need to be resolved before committing the model's state to the next tick~\cite{ptolemaeus2014system,berry1992esterel}. Therefore, execution of model(s) in a clock-domain $ \mathcal{C}_i $ at any arbitrary instance of a tick $ t\in \mathcal{T}_i $ involves resolving statuses of signals $ \mathcal{S}_i $ within $ \mathcal{C}_i $ from unknowns $ \bot $:
\begin{gather}
\frac{\zeta(\mathcal{S}_i)> 0}{\langle \mathcal{C}_i,\mathcal{S}_i \rangle \to \langle \mathcal{C}_i',\mathcal{S}_i' \rangle,\ \zeta(\mathcal{S}_i)>\zeta(\mathcal{S}_i'),\ t_{\mathcal{C}_i}=t_{\mathcal{C}_i'}} \label{eq:1} \\
\frac{\zeta(\mathcal{S}_i)= 0}{\langle \mathcal{C}_i,\mathcal{S}_i \rangle \to \mathcal{C}_i',\ t_{\mathcal{C}_i}<t_{\mathcal{C}_i'}}\label{eq:2}
\end{gather}
$\zeta: 2^{\mathcal{S}} \to \mathbb{N}$ is a function that returns the number of unresolved signals whose status is $\bot$. Intuitively speaking, Eq.~\eqref{eq:1} states that when a clock-domain makes a transition and could not resolve all signal statuses due to a cyclic relationship between them, i.e. $\zeta(\mathcal{S}_i)>\zeta(\mathcal{S}_i')$, the time for the clock-domain does not advance ($ t_{\mathcal{C}_i} = t_{\mathcal{C}_i'}$). On the other hand, when the statuses of all signals are resolved, i.e.~$ \zeta(\mathcal{S}_i)=0 $ and the clock-domain makes a macro-step transition, time progresses ($ t_{\mathcal{C}_i}<t_{\mathcal{C}_i'} $) and the number of unresolved signals in $ t_{\mathcal{C}_i'} $ again becomes greater or equals to zero. 

Micro-step transitions shown in Eq.~\eqref{eq:1} occur \textbf{\textit{within the same tick}} and they are totally ordered as in \textit{super-dense time}~\cite{ptolemaeus2014system}, which further assigns individual micro-step transitions with an index $ n_{\mu}\in\mathbb{N} $. This is needed to find constructiveness of model compositions (called fixed-point). We do not discuss on this matter in this paper and interested readers are referred to~\cite{ptolemaeus2014system} for more details.

Synchronisation points of different clock-domains are unrelated to each other where each clock-domain has its own notion of tick,
This means clock-domains are asynchronous with each other in the sense that their logical ticks $n$ are not required to be aligned in the physical timeline $r$: 
\begin{gather}
\frac{\langle \mathcal{C}_1,\mathcal{S}_{1} \rangle \to  \mathcal{C}_1' }{ \langle \mathcal{C}_1 >< \mathcal{C}_2, \mathcal{S}_{1} \rangle \to  \mathcal{C}_1'>< \mathcal{C}_2,\ t_{\mathcal{C}_1}< t_{\mathcal{C}_1'}}\\ 
\frac{\langle \mathcal{C}_2,\mathcal{S}_{2} \rangle \to \mathcal{C}'_2 }{ \langle \mathcal{C}_1 >< \mathcal{C}_2, \mathcal{S}_{2} \rangle \to  \mathcal{C}_1>< \mathcal{C}_2',\ t_{\mathcal{C}_2}<t_{\mathcal{C}_2'}}\\
\frac{\langle \mathcal{C}_1,\mathcal{S}_{1} \rangle \to   \mathcal{C}_1' ,\  \langle \mathcal{C}_2,\mathcal{S}_{2} \rangle \to \mathcal{C}'_2 }{ \langle \mathcal{C}_1 >< \mathcal{C}_2, \mathcal{S}_{1}\cup \mathcal{S}_{2} \rangle \to  \mathcal{C}_1'>< \mathcal{C}_2',\ t_{\mathcal{C}_1}< t_{\mathcal{C}_1'}\wedge t_{\mathcal{C}_2}< t_{\mathcal{C}_2'}}
\end{gather}
where $ >< $ indicates the asynchronous composition operator. Intuitively, the models at the top level of the GALS system can make state transitions in any arbitrary order given that the system is clock-domain starvation-free.

\section{Case Study: Fault Monitoring and Analysis}\label{sec:case}

\begin{figure}[t!]
	\centering
	\includegraphics[page=4,scale=0.55]{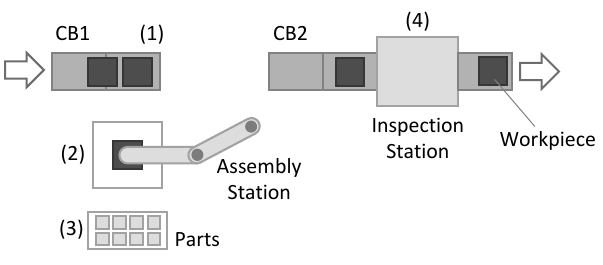}
	\caption{A graphical overview of the robot assembly and digital image inspection system}
	\label{fig:case-study}
\end{figure}

\begin{figure}[t!]
	\centering
	\includegraphics[page=5,width=0.86\linewidth]{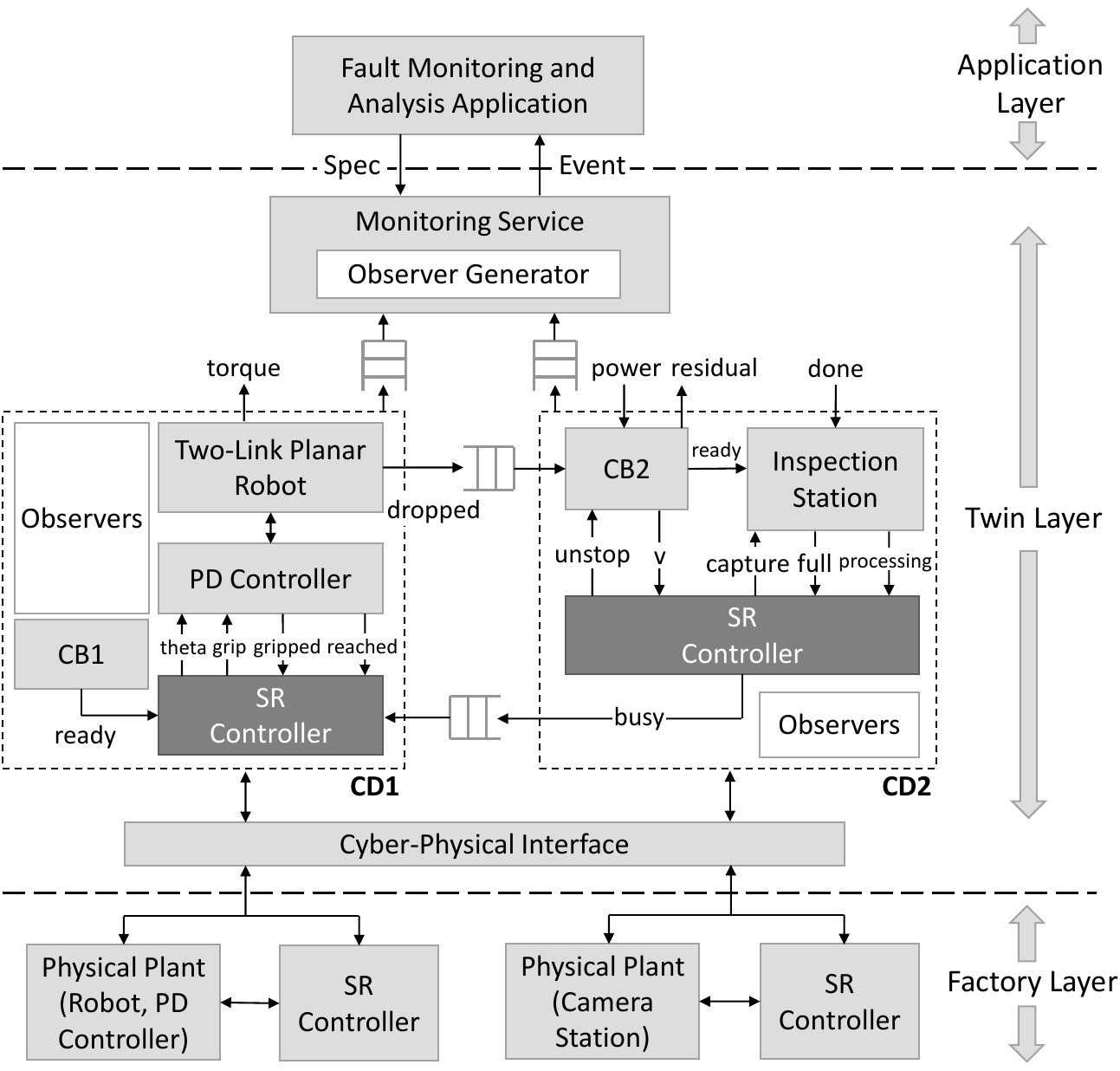}
	\caption{An architecture of the manufacturing system for fault monitoring and analysis}
	\label{fig:case-arch}
\end{figure}

In this section, a fault monitoring and analysis (FMA) application is presented as a case study for demonstrating digital twin creation and its use. The system consists of a two-axis planar robot for assembling parts of electronic products and a station for inspecting potential defects through digital image processing. A graphical illustration of the system consisting of an assembly station and a defect inspection station is shown in Fig.~\ref{fig:case-study}. The work process of this system starts with an arrival of a workpiece pallet carrying a partially finished product at (1), which triggers the two-axis robot to pick and place the item on the assembly area at (2). The robot then starts assembling the product using the parts from (3). After the job is done, the robot transfers the product on the following conveyor belt so that the item can be examined by the inspection station at (4). The inspection takes a finite amount of time to complete. Upon completion of the inspection process, the product continues to travel along the conveyor belt to the next workstation for further processing. The three-layer digital twin architecture of this system is shown in Fig.~\ref{fig:case-arch}. The Factory Layer contains two pairs of physical plants and controllers. The first pair consists of a two-axis robot integrated with a proportional-derivative (PD) controller, and an SR controller that controls a reference point fed to the PD controller. The other pair consists of a camera module and an SR controller that triggers capturing an image upon the arrival of a product at the inspection station.

\subsection{Digital Twin Models}

Twin Layer shown in Fig.~\ref{fig:case-arch} utilises heterogeneous models of four different types -- (1) the mathematical model of the two-link planar robot, the PD controller, and the dynamics of the conveyor belts captured using Modelica language~\cite{fritzson2010principles} exported as FMUs, (2) states of the camera station captured in a Petri-Net, (3) reactive control-logic in Esterel, and (4) finite automata as state observers.

\subsubsection{Two-Link Planar Robot}

The dynamics of the two-link planar robot is derived using Lagrange $ L $, which is the difference between the kinetic and potential energy~\cite{murray2017mathematical}.
\begin{equation}
L(q,\dot{q})=T(q,\dot{q})- V(q)
\end{equation}
where $ q $ is a vector of generalised coordinates of the system which in this case are joint angles of the rigid bodies. Then the Lagrange equation describing the dynamics of the robot is:
\begin{equation}
\frac{d}{dt}\frac{\partial L}{\partial\dot{q}}-\frac{\partial L}{\partial q}=F
\end{equation}
where $ F $ is an external force (i.e. torque $ \tau $) acting on the joints. Assuming the movement of the robot is restricted to an $ xy $ plane, the motion of the robot can be derived as follows.
\begin{equation}\label{eq:dyn-robot}
B(\theta)\begin{bmatrix}
\ddot{\theta}_1 \\
\ddot{\theta}_2
\end{bmatrix} + C(\dot{\theta},\theta) = F
\end{equation}
where the first term represents the inertial forces, and the second term represents the Coriolis and centrifugal forces. $ \theta_1 $ and $ \theta_2 $ are the joint angles for the first and the second link, respectively. Matrices $ B $ and $ C $ are defined as
\begin{gather*}
B=\begin{bmatrix}
\begin{smallmatrix}
\mathcal{I}_{1} + \mathcal{I}_{2} + m_1r^2_1 +m_2(l^2_1+r_2^2)+\\ 2m_2l_1r_2cos\theta_2 
\end{smallmatrix}
& \begin{smallmatrix} \mathcal{I}_{2}+m_2r^2_2 + m_2l_1r_2 cos\theta_2\end{smallmatrix}\\[6pt]
\mathcal{I}_{2}+m_2r^2_2 + m_2l_1r_2 cos\theta_2 & \mathcal{I}_{2}+m_2r^2_2
\end{bmatrix}\\
C=\begin{bmatrix}
-m_2l_1r_2sin\theta_2\dot{\theta}_2 & -m_2l_1r_2sin\theta_2(\dot{\theta}_1+\dot{\theta}_2) \\m_2l_1r_2sin\theta_2\dot{\theta}_1 & 0
\end{bmatrix}
\begin{bmatrix}
\dot{\theta}_1\\
\dot{\theta}_2
\end{bmatrix}
\end{gather*}
where $ m_i $, $ l_i $, and $ r_i $ are mass, length, and distance to the centre of mass for the $ i $'th link, respectively. $ \mathcal{I}_{i} $ is the moment of inertia perpendicular to the $ xy $ plane and relative to the $ i $'th frame at the centre of mass of the link. The structure of PD control is
\begin{equation}\label{eq:pd}
\begin{bmatrix}
	u_1 \\ u_2
\end{bmatrix} =
\begin{bmatrix}
K_{p1}(\theta_{1f} - \theta_1) - K_{D1}\dot{\theta}_1\\
K_{p2}(\theta_{2f} - \theta_1) - K_{D2}\dot{\theta}_2
\end{bmatrix}
\end{equation}
where $ \theta_{1f} $ and $ \theta_{2f} $ are desired joint angles for each link. Then combining Eq.~\eqref{eq:pd} and \eqref{eq:dyn-robot} after rearrangement we get
\begin{equation}
\begin{bmatrix}
\ddot{\theta}_1 \\ \ddot{\theta}_2
\end{bmatrix}=
B(q)^{-1}[-C(\dot{q},q)] +
\begin{bmatrix}
K_{p1}(\theta_{1f} - \theta_1) - K_{D1}\dot{\theta}_1\\
K_{p2}(\theta_{2f} - \theta_2) - K_{D2}\dot{\theta}_2
\end{bmatrix}
\end{equation}
A Modelica model has been developed for this model whose reference angle $ [\theta_{1f},\theta_{2f}]^T $ is controlled by an SR program developed in Esterel.

\subsubsection{The Camera Station}

The camera station is implemented using a Petri-Net as shown in Fig.~\ref{fig:cam}, which consists of transitions (red coloured boxes), places (blue coloured circles), and arcs (arrows). In a Petri-net, the transitions fire when all of their input places, which have an outgoing arc into the transitions, contain at least one token (shown as a black dot). When transitions fire, they consume tokens from their input places and produce tokens in output places. Arcs that no parent are mapped as input signals $ (s_{st},s_v) $ (Definition~\ref{def:dt}) as shown in Fig.~\ref{fig:case-arch}, which are connected to the neighbouring models within the same clock-domain. State of the Petri-Net is captured based on the number of tokens in each place. The algorithm of the inspection process is not directly captured in this model and only the event $ done $ is to be received from the physical plant upon its completion. Therefore, the model of this digital twin is coarser than the model of the two-link planar robot. Such a choice can be made when the application does not require a detailed view of the inspection process through the twin, which is desirable for achieving scalability.

The camera station Petri-Net has an initial marking with a single token in the place $ next $. When a status of a signal $ incoming $ changes from 0 to 1 in any of the ticks, i.e. $ s_{st}=1 $ in $ t_{i} $ and $ s_{st}=0 $ in $ t_{i-1} $, the transition fires and the number of tokens in the $ queue $ increases by one. $ ready $ indicates a detection of the workpiece at the inspection station via an infra-red sensor on the conveyor belt (\texttt{CB2} in Fig.~\ref{fig:case-arch}), which fires the connected transition and creates one token each in the places $ inprocess $ and $ precapture $ simultaneously. Start of the inspection process is initiated by the SR controller via signal $ capture $ and its completion by the physical plant via $ done $. This process repeats until no tokens (workpieces) remain in the $ queue $. Outputs of this model $ full $ and $ processing $ are set to 1 when the number of workpieces in the queue is $ > 3 $ and the inspection is in-process, i.e., there is a token in $ inprocess $. Firing transitions are done at each clock-domain tick following the SR MoC. The model is developed and compiled into executable code using the IOPT-Tools~\cite{gomes2013iopt}, which is freely available to use online at \texttt{http://gres.uninova.pt/IOPT-Tools/login.php}.

\begin{figure}[t!]
	\centering
	\scalebox{0.77}{
	\begin{tikzpicture}[
	every place/.style={draw=blue, fill=blue!20},
	every transition/.style={draw=red, fill=red!20},
	every node/.style={align=center},
	]
	\node[transition] (t1) {};
	\node[left=5mm of t1] (incoming) {$incoming$};
	\node[place,right=of t1,label=above:{$queue$\\\scriptsize\color{blue}$full=queue.token > 3$}] (p1) {};
	\node[transition,right=of p1] (t2) {};
	\node[above=5mm of t2] (ready) {$ready$};
	\node[place,below=of t2,label={[align=left]right:{$inprocess$\\\scriptsize\vspace{-5pt}\color{blue}$processing=$\\\color{blue}\scriptsize$inprocess.token > 0$}}] (p2) {};
	\node[place,right=1.1cm of t2,label=above:$precapture$] (p3) {};
	\node[transition,below right=of p3] (t3) {};
	\node[above=5mm of t3] (capture) {$capture$};
	\node[place,below=of t3,label=above left:$wait$] (p4) {};
	\node[transition,below=25.5pt of p2] (t4) {};
	\node[place,left =of p2,label=left:$next$] (p5) {} node[token] at (p5) {};
	\node[left=5mm of t4] (done) {$done$};
	\node[align=left,below left=5pt and -60pt of incoming] (legend) {
		\renewcommand{\arraystretch}{0.9} 
		\noindent
		\footnotesize
		\begin{tabular}{l}
		\toprule
		\multicolumn{1}{c}{Inputs} \\
		\midrule
		$ incoming $  \\
		$ ready $  \\
		$ capture $ \\
		$ done $ \\
		\bottomrule
		\end{tabular}
		\\[2pt]
		\footnotesize
		\renewcommand{\arraystretch}{0.9} 
		\begin{tabular}{l}
		\toprule
		\multicolumn{1}{c}{Outputs} \\
		\midrule
		$ full $  \\
		$ processing $  \\
		\bottomrule
		\end{tabular}
	};
	
	\draw (t1) edge[post] (p1);
	\draw (p1) edge[post] (t2);
	\draw (incoming) edge[post] (t1);
	\draw (ready) edge[post] (t2);
	\draw (t2) edge[post] (p2) edge[post] (p3)  edge[pre] (p5);
	\draw (p3) edge[post] (t3);
	\draw (t3) edge[post] (p4) edge[pre] (capture);
	\draw (t4) edge[pre] (p2) edge[pre] (p4) edge[post] (p5) edge[pre] (done);
	\end{tikzpicture}
}
	\caption{State transitions of the inspection station captured in a Petri-Net}
	\label{fig:cam}
\end{figure}

\subsubsection{Conveyor Belt}

Movement of the workpiece pallets on the conveyor belts is modelled using velocity that varies depending on the instantaneous power of the motor described as follows~\cite{liu2004online}:
\begin{align}\label{eq:cbp}
p(t) =&\frac{\sqrt{3}}{2}\Big\{U_mI_m[cos(2\omega_1t-\varphi)+cos\varphi]+\nonumber \\
&\sum^{\infty}_{m=1}\{U_mI_{ec1}[cos((2\omega_1-m\omega_r)t-\varphi_{ec1})+\nonumber\\
&\hphantom{\sum^{\infty}_{m=1}}cos(m\omega_rt+\varpi_{ec1})]+U_mI_{ec2}[cos((2\omega_1+m\omega_r)t\nonumber\\
&\hphantom{\sum^{\infty}_{m=1}\{}-\varphi_{ec2})+cos(m\omega_rt-\varphi_{ec2})]\}\Big\}
\end{align}
where $ U_m $ and $ I_m $ are the maximum three phase line-to-line voltage and supply current, respectively, $ \omega_1 $ is the supply angular frequency, $ \varphi $ is the phase angle, $ m $ is some integer value, and $ I_{ec1}$, $ I_{ec2} $ and $ \varphi_{ec1} $, $ \varphi_{ec2} $ are describing the induced current and corresponding phase angle due to the motor fault, which will be described in Section~\ref{sec:use}.

\subsubsection{Synchronous Reactive (SR) Controllers}

The reactive controllers shown as dark rectangles in Fig.~\ref{fig:case-arch} that control the plant models (light-grey rectangles) are developed using Esterel language. Esterel follows the SR MoC and a complete list of kernel statements and formal semantics can be found in~\cite{berry1992esterel}. We omit the detailed implementation of this control logic due to lack of space.

\subsection{Use-Case Scenario}\label{sec:use}

The twin models grouped into the same clock-domain, are interconnected and executed based on the SR MoC where inputs and outputs of the models are synchronised and resolved at tick boundaries, i.e. BOT and EOT. In this scenario, we use the delayed signal communication model, where all the signal events generated from the models are delayed by one tick, to avoid the causality problem~\cite{ptolemaeus2014system} in the SR MoC. From a digital system designer's perspective, this is equivalent to having a register on a component's input port such that the input is delayed by a single clock tick. Communication between models in two different clock-domains is done via point-to-point FIFO buffers whose sizes are predefined to prevent buffer overflows.

Online data of the physical plant are generated via a set of simulations in Modelica~\cite{fritzson2010principles}. A digital twin is executed in parallel with the physical plant, which is emulated by feeding the online data to the twin models. The role of the FMA application is to find the root cause of abnormal activities in the physical system via the digital twin by utilising both the online data and the models of the twin. In this section, detection of two different types of faults is illustrated based on the workpiece inspection time. 

\begin{enumerate}
	\item \textit{A fault in the conveyor belt's actuator}. In this scenario, a fault is indicated by intermittent spikes in the instantaneous power spectrum of the induction motor due to the induced motor stator current from the unbalanced rotor~\cite{liu2004online}.
	\item \textit{A fault in the actuator of the two-link planar robot}. A control algorithm for the workpiece placement on the conveyor belt \texttt{CB2} is disturbed by actuator fault in the robot manipulator. This results in the misalignment of workpieces that causes increase in delays in the inspection process. We assume \textit{ramp} actuator fault in this scenario where the manipulator's joints gradually drift from the actual torque reference set by the controller~\cite{mcintyre2005fault}.
\end{enumerate}

The start and the end time of the inspection process are monitored by the FMA application. The monitoring properties (specifications) requested by the application during runtime, are transformed into a set of observers (see the monitoring service in \mbox{Fig.~\ref{fig:case-arch}}). In this work, we use the fragment of Linear Temporal Logic (LTL) called co-safe LTL~\cite{bhatia2010sampling}, which can be translated into finite automata for monitoring certain events generated from the digital twin model.

\subsubsection{Using the Model and Data Online}

\begin{figure*}[!t]
	\pgfplotsset{every tick label/.append style={font=\scriptsize},
				 every axis label/.append style={font=\footnotesize}}
	\centering
	\subfloat[]{
		\includegraphics[]{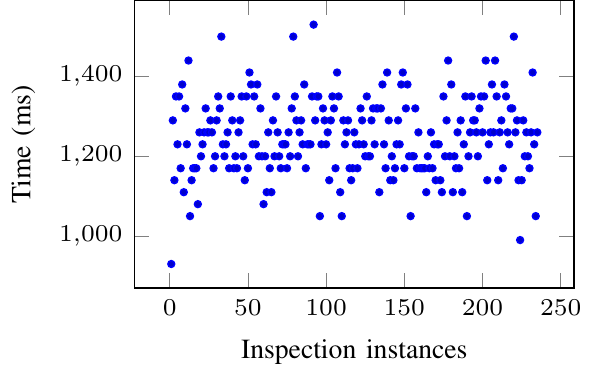}
		\label{fig:sketch}}%
	\subfloat[]{
		\includegraphics[]{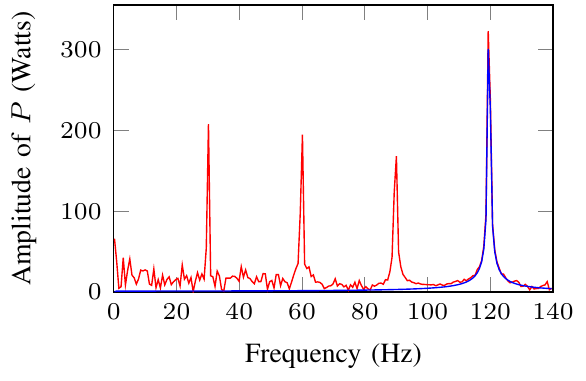}
		\label{fig:sketch-4}}%
	\subfloat[]{
		\includegraphics[]{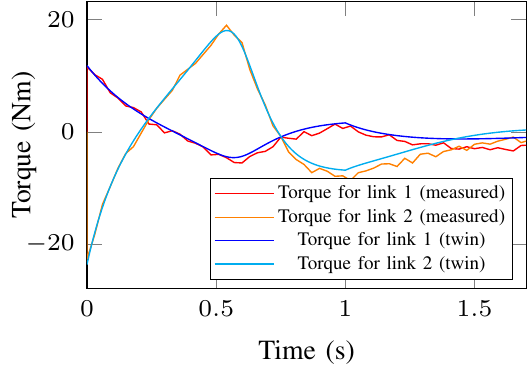}
		\label{fig:torque}}
	\caption{Results of the case study: \protect\subref{fig:sketch} process times for inspection station under normal condition, \protect\subref{fig:sketch-4} instantaneous power $ P $ spectrum for the conveyor belt's motor when it is in normal condition (blue line) and with eccentricity (red line) due to unbalanced rotor and \protect\subref{fig:torque} difference in torque generation due to ramp actuator fault for the two-link planar robot.}
	\label{fig:results}
\end{figure*}

Online usage of model and data in this case-study are twofold. First, the digital twin models are synchronised with the physical system by incorporating sensor data in the executions of the models. For example, the completion of the inspection process, indicated by a generation of an event $ done $ from the physical machine, triggers the transition in the Petri-Net model of the inspection station to fire. On the other hand, feedback from the two-link planar robot is used in a closed-loop setting for tracking the manipulator's trajectory. Second, the model is executed in parallel with the physical system, and the online data and the model simulation data are combined to generate a residual value for detecting discrepancies. 

The fault classification process from the inspection delay is described as follows. The FMA application first detects an abnormal increase in the inspection process via observers which is measured by the time between the occurrences of binary signals $ capture $ and $ inprocess $, which are shown in Fig.~\ref{fig:cam}. In this example, the twin with the online data is executed to collect a total of 235 instances of inspection times. Fig.~\ref{fig:sketch} shows inspection times under a normal operating condition where all instances fall in between $ [990,1530] $ with an average of 1,200 milliseconds. In the case of the actuator faults from the conveyor belt \texttt{CB2} or the two-link planar robot, we assume intermittent increases in the inspection time indicate a need for further inspection of the physical system is required via the digital twin models. In such a case, possible sources of the fault are first located via causal correlations between the failure node (inspection station) and the causal node (the conveyor belt or the two-link planar robot) based on their input and output signals and queue interconnections. To confirm the fault, the causal nodes are examined by the FMA application via a residual value generated from the online digital twin simulation and the sensory data gathered from the physical plant. In the case of conveyor belt motor fault, the fault characteristic components that appear in the instantaneous power spectrum at frequencies 30, 60, and 90 Hz are compared with the normal component frequency of 120 Hz as shown in Fig.~\ref{fig:sketch-4}. This is obtained by computing the fast Fourier transform of the instantaneous power (Eq.~\ref{eq:cbp}). On the other hand, a gradual drift of the robot manipulator's joint from the reference torque generated from its digital twin model is examined in the case of the ramp actuator fault as shown in Fig.~\ref{fig:torque}.

The anomaly detection algorithm for comparing the online data with the data generated from the digital twin model can be as simple as checking a threshold crossing using the model as reference points or other approaches such as $ \chi^2 $ test, k-nearest neighbour and Bayesian methods. Once the residual value is analysed by the FMA application, it performs recovery or preventive control actions accordingly through the control services in the Twin Layer as shown in Fig.~\ref{fig:title}.

\subsection{Model Fidelity}

\begin{figure}[t!]
\tikzset{rounded/.style={draw,rectangle,rounded corners}}
\centering
\begin{tikzpicture}[
	node distance=2cm,auto,
	initial text=,
	every node/.append style={align=center,font=\footnotesize},
	]
	\node[rounded,initial above] (idle) {$idle$};
	\node[rounded] (conv) [right=of idle] 
	            {$conv$\\$\theta_1=f_{c1}(t)$\\$ \theta_2=f_{c2}(t) $};
	\node[rounded,align=center] (asm) [left=of idle] 
	            {$asm$\\$\theta_1=f_{a1}(t)$\\$ \theta_2=f_{a2}(t) $};
	
	\path[-{>[length=3]}] (idle) edge [bend left] node[above right, xshift=-0.5cm] 
						  {$ move_1(\theta_{1f},\theta_{2f}) $} (conv)
					 edge [bend left] node[below left, xshift=0.5cm] 
				  	      {$ move_2(\theta_{1f},\theta_{2f}) $} (asm)
			  (conv) edge [bend left] node[below right, xshift=-1cm]
			              {$ reached_1(\theta_{1f},\theta_{2f},\theta_{1},\theta_{2}) $} (idle)
			  (asm) edge [bend left] node[above left, xshift=1cm]
			  			  {$ reached_2(\theta_{1f},\theta_{2f},\theta_{1},\theta_{2}) $} (idle)
	;
\end{tikzpicture}
	\caption{A finite state machine model for the two-link planar robot using linear interpolation}
	\label{fig:robot-fsm}
\end{figure}
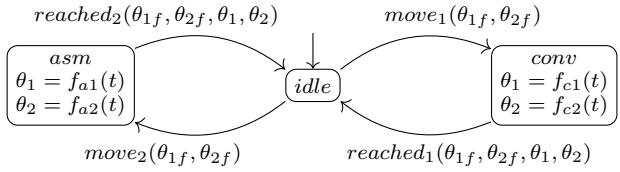

\begin{figure}[t!]
	\centering
	\includegraphics[width=\columnwidth]{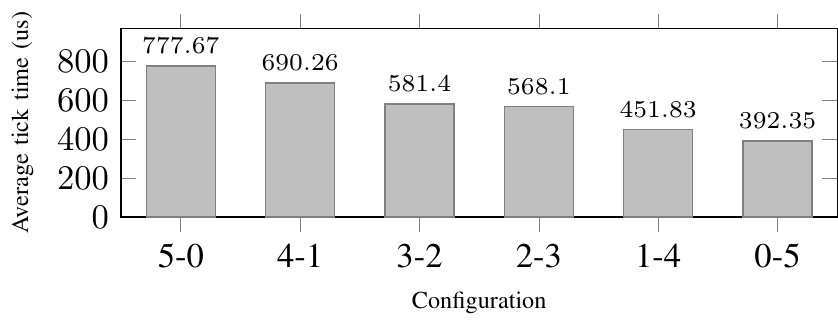}
	\caption{Average execution times for 5 FMA applications with high and low (h-l) model fidelity configurations}
	\label{fig:results-dm}
\end{figure}

An effect of low and high fidelity models is investigated by utilising heterogeneous models for the FMA case study. In this experiment, the FMA application shown in Fig.~\ref{fig:case-arch} is duplicated for five times and an execution time of the digital twin models is measured with varying number of low and high fidelity models. More specifically, two-link planar robot models are replaced with a state machine-based low fidelity model as shown in Fig.~\ref{fig:robot-fsm}. This model captures the movements of the robot's manipulator between between the assembly station and the second conveyor belt as depicted as (2) and \texttt{CB2} in Fig.~\ref{fig:case-arch}. Two states labelled as $ asm $ and $ conv $ indicate the movement of the manipulator's end effector from the conveyor belt to the assembly station and vice-versa. In this model, joint angles for the manipulator's links are pre-computed and stored in tables. Functions $ f_x $ in $ asm $ and $ conv $ computes linear interpolation between the pre-computed values based on the current time $ t $. In this experiment, the twin models are all grouped into a single clock-domain whose average tick time is measured. The results are shown in Fig.~\ref{fig:results-dm} where the notation h-l is used to indicate the ratio of the high and low fidelity models. The graph shows improvements in the average tick time of the digital twin model up to 1.98 times as the robot models are reduced to the finite state machine model. The result indicates the heterogeneous models with varying fidelities are desired features for building a digital twin that requires real-time simulation capability unlike traditional offline simulation tools.

\section{Conclusions and Future Work}\label{sec:conclusion}

We presented a digital twin architecture TiLA for building a digital twin based on Globally Asynchronous Locally Synchronous (GALS) model of computation (MoC). Digital twin in TiLA can be created from heterogeneous models with minimal support for a predefined interface that enables model composition and execution in GALS. In addition, TiLA supports a co-simulation capability through FMU and implements the GALS-based master scheduling algorithm.

The current FMI standard lacks support for some of the mixed discrete event-based phenomena, which are present in the synchronous reactive subset of GALS MoC. For future work, the inclusion of the upcoming FMI 3.0 standard that enhances \textit{hybrid co-simulation} capability in the current GALS semantics would be advantageous. Support for more expressive temporal logics such as Signal Temporal Logic~\cite{maler2004monitoring} and its variants could also be used for both the monitoring specification and model synthesis. 

%

\bibliographystyle{IEEETran}
\bibliography{bibliography}

\end{document}